# Design and Performance Analysis of a Chemically-Etched Flexible NFC Tag Antenna

Muhammad Enayetur Rahman[1], Marwan Abdelatti[2], Manbir Singh Sodhi[2], Kunal Mankodiya[1]
Wearable Biosensing Lab, Dept of Electrical, Computer, and Biomedical Engineering, University of Rhode Island, USA
Dept of Mechanical, Industrial and Systems Engineering, University of Rhode Island, USA
menayet@uri.edu

*Abstract*—**Near Field Communication (NFC) is a perfect example of ubiquitous computing that is secured, short-ranged, low-powered contactless communication. High demand is predicted for NFC, especially with wearables, and the variety of applications may require that this technology be fabricated onto different materials. In this research, we first designed flexible NFC antennas based on mathematical models. A simulation of the NFC coil inductance was performed and verified using the predictive models. Later, two flexible NFC antennas, including 160x80 mm$^2$ (rectangular) and 80x80 mm$^2$ (square) were fabricated through a chemical etching process and verified at 13.56 MHz frequency. Two experiments 1) Read range detection and 2) User experience study on 22 participants validate longer read range and shorter connection time of our fabricated flexible NFC tags.**

## I. INTRODUCTION

Near-field communication (NFC) is a radio-frequency identification system (RFID) that enables fast and secure communication between two devices over a short range using the 13.56 MHz RFID band [1]. This two-way interaction enables people to access coupons in different outlets, or even pairing two NFC-enabled devices with just a single tap [2]. NFC technology benefits from the Internet of Things (IoT) to expand rapidly as most of the smartphones have NFC readers [3, 4]. Since NFC is a quick and easy way of obtaining data by placing the tag near to the reader without any pairing mechanism, it broadens the low-cost NFC sensors market.

In 2018, ISO 15693 has amended the readable range of NFC to 0.944 m [5]. In order to achieve the maximum energy from an NFC reader, material properties should be considered in the design of a flexible NFC (f-NFC) tag antenna. As a flexible NFC (f-NFC) tag can be developed using polymer-metal composites, the interconnect conductivity plays a key role in the maximum achievable NFC detection read range. Because of its high resistivity, the graphitic carbon tag does not resonate at 13.56 MHz and hence, is not suitable for NFC tag application [6]. The study showed that, as the tag resistance increases, the read range decreases dramatically. For a flexible NFC (f-NFC) antenna, the research challenge is to design it on flexible/soft materials, fabricate and integrate it with an NFC tag chip and accurately tune it at the right frequency (13.56 MHz [5]).

This research is focused on designing, verification and user centric testing of a flexible NFC (f-NFC) antenna on its detectable read range. We developed, tested and verified a flexible NFC (f-NFC) tag antenna considering different parameters on flexible fiberglass material. The f-NFC antenna parameters include shapes, sizes, and conductor widths. We exploited several of these parameters in our design and investigated their effects on the read range performance of the f-NFC antennas by mathematical analysis, simulation, and empirical testing. Finally, we let users test our developed f-NFC tags using their smartphones and submit their feedback.

## II. BACKGROUND

### A. Components of an NFC system

*1) NFC reader:* An NFC reader is capable of exchanging information with another NFC device in both directions. NFC readers that are integrated into smartphones will be our focus in this paper. NFC integrated smartphones enhance the Internet of Things (IoT) world [7].

*2) NFC tag:* Usually, an NFC tag does not have a power source. Instead, they receive the power from the Radio Frequency (RF) signals sent from the NFC reader. A small amount of information can be stored in the memory available on the NFC tag. The memory capacity varies depending on the tag IC chip.

### B. Flexible NFC (f-NFC) Antennas

Flexible NFC (f-NFC) antennas have excellent potential for usage in wearable applications in the areas of healthcare, sports, consumer electronics, and many other sectors. Despite technical and design challenges, f-NFC has been advancing rapidly and can be categorized into two groups:

*1) Thin-film f-NFC:* Thin-film wearable antennas involve metal deposition upon flexible and thin substrates. This substrate could be fiberglass, Kapton, Polyethylene terephthalate (PET), paper, or polydimethylsiloxane (PDMS) [8]. Flexibility, light-weight, low-cost, and robustness are the characteristic factors for the thin-film f-NFC. Procedures to fabricate these types of antennas include screen printing, sputtering, manual deposition of conductive polymers, photolithography injection of liquid metal alloy, and inkjet printing.

*2) Fabric f-NFC:* In this type, a conductive thread is usually woven or deposited onto the fabric substrate. Some examples of fabric substrates are Cordura, felt, Lycra, Cotton, Polyester, and Polycot [9, 10]. These types of fabrics have low permittivity and tangent values [8]. Fabrication techniques could be, adhesive [11], screen printing [12], inkjet printing [13], embroidery of conductive threads [14].

This research is supported by the RI Innovation Voucher Grant.

## III. MATERIALS & METHODS

We designed and implemented a thin-film chemically-etched f-NFC antenna connected to an NFC tag IC: *NTAG I²C Plus* manufactured by NXP® Semiconductors.

### A. NFC Tag IC:

An NFC tag IC of type NTAG I²C tag was selected for the experiment due to two main features: 1) NTAG is very secure, low priced, and also capable of harvesting energy to a microcontroller. 2) NTAG data transmission speed is very high (106 kbps) compared to ICODE type ICs, which is 26.5 kbps [15].

### B. Simulation - Coil Inductance Calculations:

Fig. 1 shows the equivalent circuit of the NFC tag. The IC chip is represented by a resistance $R_c$ and a capacitance $C_c$ connected in parallel [15]. The equivalent antenna circuit is represented by serially connected resistance $R_{s-ant}$ and a coil $L_{s-ant}$. The capacitor $C_{tune}$ is an external tuning capacitor to help reach a resonance frequency of 13.56 MHz. An open circuit voltage is delivered by the antenna which depends on the electromagnetic field strength and the antenna parameters. The inductance of a square antenna [16, 17] is calculated by using Modified Wheeler's formula:

$$L_{ant} = K_1 \mu_0 N^2 \frac{d}{1+ K_2 P} \quad (1)$$

Where $\mu_0$ is the permeability of vacuum, $N$ is the number of turns, $d$ is the mean coil diameter, $d = (d_{out} + d_{in})/2$ in mm, $d_{out}$ = outer diameter, $d_{in}$ = inner diameter, $p = (d_{out} - d_{in})/(d_{out} + d_{in})$ in mm. $K_1$ and $K_2$ are constants that depend on the antenna layout with values: 2.34 and 2.75, respectively. The critical parameter in the design of an NFC coil is the equivalent inductance $L_{ant}$ of the antenna at 13.56 MHz. The stray capacitance between the turns is usually in the order of a few pico-farads and can be neglected [18].

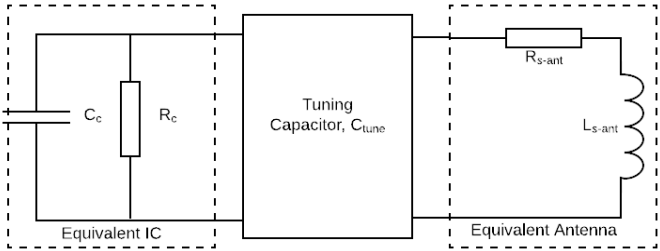

Fig. 1. NFC tag equivalent circuit

The eDesignSuite from STMicroelectronics was used for the design according to the parameters shown in Table I.

TABLE I. DESIGN PARAMETERS OF F-NFC

| Parameter Type | Design Parameters | Antenna 1 | Antenna 2 |
|---|---|---|---|
| Antenna Geometry | Length (mm) | 160 | 80 |
| | Width (mm) | 80 | 80 |
| | Number of turns | 4 | 3 |
| Conductor | Conductor width (mm) | 0.5 | 0.6 |
| | Spacing between turns (mm) | 2 | 2 |
| | Thickness (mm) | 0.0175 | 0.0175 |
| Substrate | Thickness (mm) | 0.127 | 0.127 |
| | Permittivity | 4.6 | 4.6 |

### C. Flexible NFC (f-NFC) Fabrication:

A chemical etching process was used to fabricate the antennas on flexible, single-sided, fiberglass PCB boards.

The antennas were first designed according to the specifications defined in Table I by Eagle software then the layout was printed on paper by a laserjet printer. Thermal transfer is used to transfer the layouts to the PCB boards which are then etched using Ferric chloride (FeCl₃) etchant solution.

The two fabricated antennas are shown in Fig. 2.

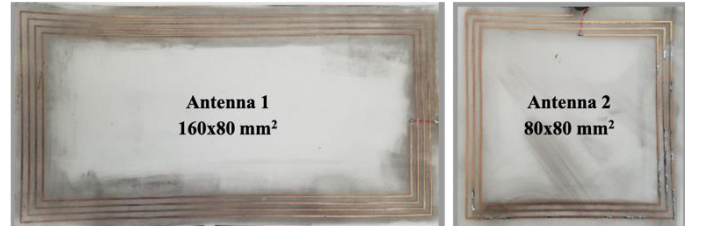

Fig. 2. Fabricated f-NFC antennas.

## IV. FLEXIBLE NFC TAG VERIFICATION

After fabricating the antennas, the equivalent series circuit $R_{s-ant}$ and $L_{s-ant}$ are measured using an impedance analyzer (HP 4294A). The values are listed in Table II. The tuning capacitor value $C_{tune}$ was calculated using the following equation:

$$f_r = \frac{1}{2\pi\sqrt{L_{s-ant} C_{eq}}} \quad (2)$$

Where the IC capacitance value $C_c$ is 50 pF [15], assuming that the antenna capacitance and the equivalent parasitic connection capacitance are negligible, $f_r$ is the resonance frequency, and $C_{eq}$ is the equivalent capacitance of $C_c$ and $C_{tune}$. The tuning capacitor $C_{tune}$ is connected either in series or in parallel according to the calculations. The Surface Mount Devices (SMD): tag IC and $C_{tune}$ are soldered to the circuit forming a complete f-NFC tag. The tag resonating frequency is measured by the impedance analyzer and also verified by a network analyzer (Agilent N3383A). Table II shows the component values where antennas resonate very close to 13.56 MHz.

TABLE II. VERIFICATION OF F-NFC ANTENNAS

| Antenna | $L_{s-ant}$ (μH) | $R_{s-ant}$ (Ohm) | $C_C$ (pF) | $C_{tune}$ (pF) | $C_{tune}$ Connection | Freq (MHz) |
|---|---|---|---|---|---|---|
| Antenna 1 | 4.85 | 10 | 50 | 56 | Series | 14.06 |
| Antenna 2 | 1.62 | 2 | 50 | 30 | Parallel | 13.98 |

## V. EXPERIMENTS

### A. Sensing Range Experiment with Different Phones:

We tested and recorded the reading range of 30 different commercial NFC tags. We designed a specialized frame, shown in Fig. 3, enabling the phone to move vertically against the f-NFC tag placed on the ground plane.

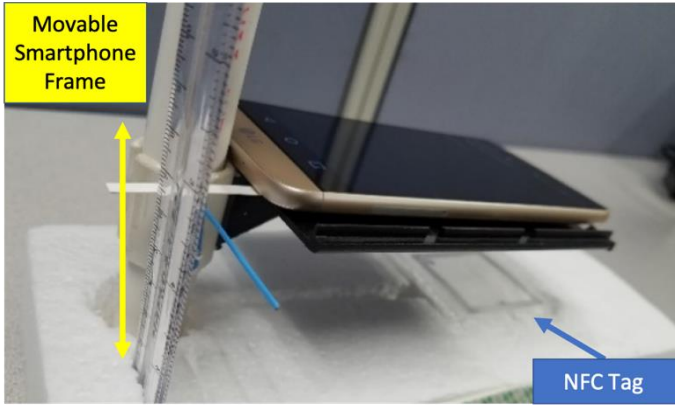

Fig. 3. NFC tag read-range measurement method

Table III shows the comparison between our tags and the best performing commercial one using the same standard (ISO-14443-3A). The tag with the large antenna achieved 8.1 cm while the smaller antenna tag achieved a range of 5 cm. The best commercial tag read range was found 5.6 cm.

TABLE III. ANTENNA READ RANGE RESULT

| Antenna | Range (cm) |
|---|---|
| Antenna 1 | 8.1 |
| Antenna 2 | 5.0 |
| Commercial tag | 5.6 |

*B. User Experience Study:*

In this stage, we aim to test the f-NFC tags with end-users. We designed two identical posters: poster A contains the commercial tag mentioned in Table III, and poster B contains our f-NFC tag (antenna 1 – 160x80 mm$^2$) since it performs better than antenna 2. We asked 22 users to interact with the two posters using their smartphones. Fig. 4. describes the boxplots of the two poster detection timings. This was a blind study. The users were not made aware of the significance of the poster labels or tags.

A smartphone application was developed to measure the duration to detect an NFC poster successfully. Finally, users were asked to fill out a survey that summarizes their experience. Fig. 4. shows poster B with f-NFC requires much less time to detect successfully from a smartphone than the best available commercial tag. The average NFC detection time of the commercial tag was 4.14 seconds, whereas f-NFC took an average time of 2.05 seconds to connect. It can be seen from the boxplot, 75% of users using poster B with f-NFC took only 0.9 to 2.115 (Quartile 3) seconds. On the other hand, 75% of poster A users took 1.37 to 6.22 (Quartile 3) seconds.

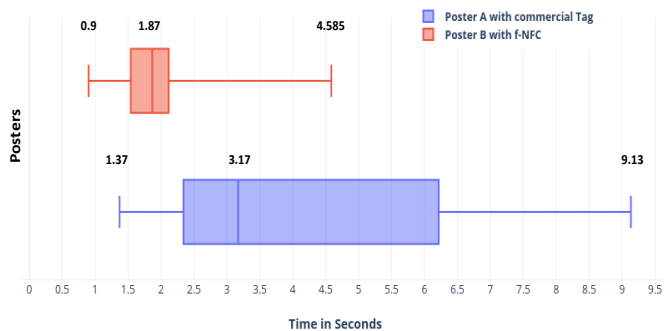

Fig. 4. Boxplot of poster detection timing

The median values of poster A and B are 3.17 and 1.87 seconds respectively.

Fig. 5 shows the users' rating (on a scale of 0 to 10) on the overall performance of both tag posters. Our fabricated f-NFC tag (poster B) received a distinctly higher rating compared to the commercial counterpart. Based on the survey, users were more satisfied with our f-NFC tag than the commercial one. 95.5% of the users had experienced a longer detection range for poster B. 63.6% of the users believed that poster B had a quicker response.

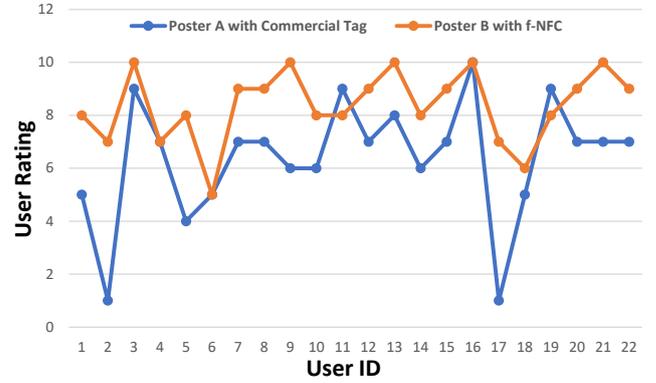

Fig. 5. User rating for posters A and B.

## VI. CONCLUSIONS

The work presented here investigated how the design parameters of a flexible antenna impact the performance of the f-NFC tag in terms of its detectable range. Two f-NFC tags were fabricated through the chemical etching process and tested for their resonant frequency, which remained close to 13.56 MHz. Our experiments revealed that the large rectangular f-NFC offered the readable range of 8.1 cm (compared to 5.6 cm - range of the commercial tag) and was connected in 2.05 seconds (compared to 4.25 seconds – response time of the commercial tag). Our f-NFC tag was rated higher in terms of users' satisfaction compared to most of the commercial NFC tags in the market. Our experimental results open a wide range of applications of this type of f-NFC tag, including but not limited to smart textiles and wearable technology. Our current work suggests to pursue more research and experimentation on flexible NFC antennas in terms of materials, geometry, fabrication process, end-users, and applications.

ACKNOWLEDGMENT

We are thankful to our industrial partner Cooley Groups. This research was supported by the Rhode Island Innovation Voucher Grant.